\documentclass[aps,amsmath,amssymb,superscriptaddress,prb,preprint]{revtex4}
\usepackage[dvips]{graphicx}
\usepackage{epstopdf}
\usepackage{dcolumn}
\usepackage{bm}
\usepackage{ulem}
\usepackage[dvips]{color}
\usepackage[dvips]{color} 
\usepackage{setspace}
\usepackage{graphicx,color}
\begin{document}
\title{Enhancement of Spin Pumping in $\mathrm{Y_3Fe_5O_{12}/Pt/Ni_{81}Fe_{19}}$ Trilayer Film}
\author{Daichi Hirobe}%
\email{daichi.kinken@imr.tohoku.ac.jp}
\affiliation{Institute for Materials Research, Tohoku University, Sendai 980-8577, Japan}
\author{Ryo Iguchi}
\affiliation{Institute for Materials Research, Tohoku University, Sendai 980-8577, Japan}
\author{Kazuya Ando}
\affiliation{Institute for Materials Research, Tohoku University, Sendai 980-8577, Japan}
\affiliation{Department of Applied Physics and Physico-Informatics, Keio University, Yokohama 223-8522, Japan}
\affiliation{PRESTO, Japan Science and Technology Agency, Saitama 332-0012, Japan}
\author{Yuki Shiomi}
\affiliation{Institute for Materials Research, Tohoku University, Sendai 980-8577, Japan}
\author{Eiji Saitoh}
\affiliation{Institute for Materials Research, Tohoku University, Sendai 980-8577, Japan}
\affiliation{WPI Advanced Institute for Materials Research, Tohoku University, Sendai 980-8577, Japan}
\affiliation{CREST, Japan Science and Technology Agency, Chiyoda, Tokyo 102-0075, Japan}
\affiliation{The Advanced Science Research Center, Japan Atomic Energy Agency, Tokai 319-1195, Japan}
\date{\today}
\begin{abstract}
We study spin pumping in a $\mathrm{Y_3Fe_5O_{12}}$(YIG)/Pt/$\mathrm{Ni_{81}Fe_{19}}$(Py) trilayer film by means of the inverse spin Hall effect (ISHE). When the ferromagnets are not excited simultaneously by a microwave, ISHE-induced voltage is of the opposite sign at each ferromagnetic resonance (FMR). The opposite sign is consistent with spin pumping of bilayer films. On the other hand, the voltage is of the same sign at each FMR when both the ferromagnets are excited simultaneously. Futhermore, the voltage greatly increases in magnitude. The observed voltage is unconventional; neither its sign nor magnitude can be expected from spin pumping of bilayer films. Control experiments show that the unconventional voltage is dominantly induced by spin pumping at the Py/Pt interface. Interaction between YIG and Py layers is a possible origin of the unconventional voltage.

\end{abstract}
\maketitle
\section{Introduction}
Spintronics has attracted attention because it is expected to realize new magnetic memories and computing devices by use of electronic spins.\cite{Zutic:2004vi} A necessity of spintronics is to understand the nature of a spin current, a flow of spin angular momentum. The concept of a spin current stems from spin-polarized transport in ferromagnetic metals. Ferromagnetic metals conduct a spin-dependent electric current, a charge current accompanied by a spin current. This current induces interesting phenomena such as giant magnetoresistance\cite{Fert:2008} and current-induced magnetization reversal\cite{Slonczewski:1996, Berger:1996, Kiselev:2003}.
 
The complete separation into spin and charge currents leads to the concept of a {\it pure} spin current, a flow of electronic spins with {\it no} charge currents. Spin pumping\cite{Mizukami:2002tt,Tserkovnyak:2005fr, Azevedo:2005} is a versatile method for generating a pure spin current. This method uses magnetization precession in a ferromagnet to inject a spin current into an adjacent metal. 

Ferromagnet (F)/normal metal (NM) bilayer films have offered themselves as the benchmark for the validity of spin pumping. Experimental and theoretical studies have concentrated on this structure and contributed to a solid understanding of spin pumping. Spin-current injection enhances magnetization relaxation in F and broadens ferromagnetic-resonance (FMR) line width. The injection is electrically detectable in NM as well; spin-orbit interaction in NM can convert an injected spin current into a charge current. This conversion is known as the inverse spin Hall effect\cite{Kimura:2007, Saitoh:2006, Valenzuela:2006, Seki:2008, Takahashi:2002, Ando:2008} (ISHE) and induces an electric field $\mathbf{E}_{\mathrm{ISHE}}$ that satisfies the relation\cite{Saitoh:2006}
\begin{equation}
\mathbf{E}_{\mathrm{ISHE}} \propto \mathbf{J}_{\mathrm{s}}\times\mathbf{\boldsymbol\sigma}. \label{ishe}
\end{equation}
Here, $\mathbf{J}_{\mathrm{s}}$ is the spatial direction of a spin current and $\boldsymbol\sigma$ is the spin-polarization vector. 

Spin pumping in F/NM/F trilayer films will be of further interest since new collective excitation can be induced by magnetic couplings between the ferromagnets. In fact, such excitation was verified by FMR measurements.\cite{Heinrich:2003,Tserkovnyak:2005fr} This paper presents an experimental investigation on spin pumping in a $\mathrm{Y_3Fe_5O_{12}(YIG)/Pt/Ni_{81}Fe_{19}(Py)}$ trilayer film by means of ISHE. DC voltage is found to be enhanced when both the ferromagnets are excited simultaneously. The enhanced voltage is unconventional in the sense that it cannot be explained by spin-pumping theory of F/NM bilayer films. Control experiments demonstrates that the unconventional voltage comes from spin pumping at the Py/Pt interface. We discuss the origin of the unconventional voltage and narrow down contributing factors on an experimental basis.

\section{method}
Figure 1(a) shows a schematic illustration of a $\mathrm{YIG/Pt/Py}$ trilayer film. The $\mathrm{YIG}$ (Pt) layer was 1 mm wide, 3 mm long and 50 nm (5 nm) thick. The $\mathrm{Py}$ layer was 1 mm wide, 2 mm long and 10 nm thick. We grew the $\mathrm{YIG}$ layer on a $\mathrm{Gd_{3}Ga_{5}O_{12}}$ (GGG) substrate by a metal organic decomposition method\cite{Ishibashi:2005}; we then sputtered the Pt layer on the YIG layer and finally deposited the Py layer by the electron-beam evaporation in a high vacuum. We attached two electrodes to the ends of the Pt layer to measure DC voltage.

The film was placed at the center of a $\mathrm{TE_{011}}$ microwave cavity at which a microwave magnetic field (electric field) maximizes (minimizes) itself. A frequency of the microwave was 9.44 GHz. We applied a static magnetic field perpendicular to the microwave magnetic field. Figure 1(b) defines the out-of-plane magnetic-field angle $\theta_{H}$ and the magnetization angle $\theta_{M, \mathrm{YIG}}$ ($\theta_{M, \mathrm{Py}}$) of YIG (Py). Spin pumping injects a spin current into the Pt layer, leaving the spin-polarization vector $\mathbf{\boldsymbol\sigma}$ parallel to the precession axis on temporal average [see Fig. 1(c)].\cite{Saitoh:2006} ISHE in the Pt layer converts the injected spin current into DC voltage. DC voltage and an FMR spectrum were measured at room temperature with changing static-field magnitude or direction.

\section{RESULTS and DISCUSSION}
Figure 2(a) shows spectra of the microwave absorption signal at $\theta_H$ = $\pm$90 deg. (i.e., with in-plane fields). We see FMR of Py (YIG) near $\mu_{0}H$ = 120 (300) mT. Correlating with FMR, DC voltage appears as shown in Fig. 2(b). Remember the direction of relevant vectors: $\mathbf{J}_{\mathrm{s}}\parallel \hat{\mathbf{y}}$, $\mathbf{\boldsymbol\sigma}\parallel \hat{\mathbf{z}}$, and $\mathbf{E}_{\mathrm{ISHE}} \parallel \hat{\mathbf{x}}$ [see Fig. 1(c)]. Here, $\hat{\mathbf{x}}$, $\hat{\mathbf{y}}$, and $\hat{\mathbf{z}}$ represent unit vectors in each direction. The voltage spectra are consistent with Eq. (\ref{ishe}). At $\theta_H$ = 90 deg., voltage is negative at FMR of Py while it is positive at FMR of YIG. This opposite polarity is due to the opposite directions of spin-current injection ($\mathbf{J}_{\mathrm{s}} \parallel \pm\hat{\mathbf{y}}$, $\mathbf{\boldsymbol\sigma}\parallel \hat{\mathbf{z}}$, and thus $\mathbf{E}_{\mathrm{ISHE}} \parallel \pm\hat{\mathbf{x}}$). Moreover, when the field direction is reversed, the voltage sign reverses at each FMR. This reversal is due to the opposite directions of spin polarization ($\mathbf{J}_{\mathrm{s}} \parallel \hat{\mathbf{y}}$, $\mathbf{\boldsymbol\sigma}\parallel \pm\hat{\mathbf{z}}$, and thus $\mathbf{E}_{\mathrm{ISHE}} \parallel \pm\hat{\mathbf{x}}$). These results all support that the voltage is dominated by ISHE. The replacement of Pt with Cu reveals irrelevance of galvanomagnetic effects such as the anomalous Hall effect and the planar Hall effect\cite{Chen:2013}. The replacement greatly reduces voltage. As a result, the voltage with Cu cannot reproduce the voltage with Pt, even considering a conductance difference between Pt and Cu. This reduction is consistent with the fact that ISHE is negligibly small in Cu owing to weak spin-orbit interaction.

Figure 2(c) shows the $\theta_H$ dependence of a resonance field $H_{\mathrm{FMR}}$. The result is consistent with previous studies of bilayer films.\cite{Ando:2011} In terms of $H_{\mathrm{FMR}}$, we observed little difference between the trilayer film and its corresponding YIG/Pt and Py/Pt bilayer films. Different saturation magnetization between YIG and Py results in the different $\theta_H$ dependence of $H_{\mathrm{FMR}}$ via a demagnetization field along the $y$-axis [see Fig. 1(b)].\cite{Ando:2011} This enables us to control magnetization dynamics in each layer. At $\theta_{H} = \pm90$ deg, resonance fields are different between $\mathrm{YIG}$ and $\mathrm{Py}$. Thus, while one magnetic layer is on resonance with large precessional amplitude, the other is off resonance with small precessional amplitude. Around $\theta_{H} = \pm20$ deg., on the other hand, resonance fields are nearly identical: magnetization precesses simultaneously in the two ferromagnets.

Figure 3(a) shows the magnetic-field dependence of voltage at various values of $\theta_{H}$. We first confine ourselves to voltage at FMR of Py. Voltage monotonically decreases in magnitude as $\theta_{H}$ approaches zero and changes its sign across $\theta_{H} = 0$ deg.. This behavior is consistent with Eq. (\ref{ishe}) because $\mathbf{E}_{\mathrm{ISHE}} \propto \mathbf{J}_{\mathrm{s}}\times\mathbf{\boldsymbol\sigma} \propto \sin \theta_{M, \mathrm{Py}}$ holds [see Fig. 1(b)]. To fit the voltage spectra near FMR of Py, we used a sum of symmetric and asymmetric Lorentzian functions\cite{Saitoh:2006}
\begin{equation}
V(H) = V_\mathrm{SYM}\frac{(\Delta H/2)^2}{(H-H_{\mathrm{FMR}})^2+(\Delta H/2)^2} + V_{\mathrm{ASYM}}\frac{2\Delta H/2(H-H_{\mathrm{FMR}})}{(H-H_{\mathrm{FMR}})^2+(\Delta H/2)^2}.
\end{equation}
Here, $\Delta H$ is the full width at half maximum and $V_{\mathrm{SYM}}$ ($V_{\mathrm{ASYM}}$) is the magnitude of a symmetric (asymmetric) component of voltage. $V_{\mathrm{SYM}}$ corresponds to ISHE and $V_{\mathrm{ASYM}}$ to the anomalous Hall effect in this study. Fig. 3(b) bottom shows that the $\theta_H$ dependence of $V_{\mathrm{SYM}}$ is roughly represented by a sine curve and reveals no difference between the YIG/Pt/Py trilayer film and a Py/Pt bilayer film.

We next focus on voltage at FMR of YIG for $\theta_{H} > 0$ deg. Fig. 3(a) shows that as $\theta_{H}$ decreases from 90 deg., voltage magnitude decreases and almost vanishes at 30 deg.. Notable is that voltage of the opposite sign reappears at 20 deg.. This reversal of sign cannot be explained by spin pumping in a YIG/Pt bilayer film. Application of Eq. (1) to the bilayer film yields voltage of a constant sign for $\theta_{H} > 0$ deg. because $\mathbf{E}_{\mathrm{ISHE}} \propto \mathbf{J}_{\mathrm{s}}\times\mathbf{\boldsymbol\sigma} \propto \sin \theta_{M, \mathrm{YIG}}$ holds. An unconventional feature is not limited to the sign of voltage: the upper inset in Fig. 3(a) clearly shows enhancement of voltage near $\theta_{H} = 18$ deg, where the resonance fields of YIG and Py are equal. Anomalies in sign and magnitude are summarized in Fig. 3(b) and (c). At this stage, we cannot conclude that the unconventional voltage originates in spin pumping at the YIG/Pt interface. We also need to consider spin pumping at the Py/Pt interface; YIG and Py can affect mutual dynamics, or spin pumping via magnetic interaction.

We note that the observed unconventional voltage cannot be attributed to heating effects  due to microwave absorption. The microwave absorption at FMR heats the magnetic layer and may cause a heat current flowing from the magnetic layer to the Pt layer. Eventually DC voltage is induced by thermoelectric effects such as the ordinary Nernst effect, the anomalous Nernst effect and the longitudinal spin Seebeck effect\cite{Uchida:2010} (LSSE). The DC voltage is proportional to $\bm{\mathrm{H}}\times\nabla T$ (ordinary Nernst effect) and $\bm{\mathrm{M}}\times\nabla T$ (anomalous Nernst effect and LSSE), where $T$ denotes temperature. The magnitude of $\nabla T$ is proportional to the microwave absorption at FMR $I_{\mathrm{FMR}}$; thus the DC voltage is proportional to $|\mathbf{H}|I_{\mathrm{FMR}}\sin\theta_{H}$ (ordinary Nernst effect) and $|\mathbf{M}|I_{\mathrm{FMR}}\sin\theta_{M}$ (anomalous Nernst effect and LSSE). Since $I_{\mathrm{FMR}}$ decreases as $\theta_{H}$ approaches zero, the heating effects cannot reproduce the observed $\theta_{H}$ dependence of voltage. 

Voltage anomalies appearing when $H_{\mathrm{FMR, YIG}} = H_{\mathrm{FMR, Py}}$ suggest that magnetization dynamics, or spin pumping is affected by a magnetic coupling between the ferromagnets. One of the possibilities is a dynamic exchange coupling\cite{Heinrich:2003}, a coupling via spin currents mediated by conduction electrons in an intermediate NM layer. This coupling persists on the spin-diffusion length scale in the NM layer and may exist in the YIG/Pt/Py trilayer film, because Pt has a spin diffusion length of around 10 nm. To examine a dynamic exchange coupling, we prepared a YIG/$\mathrm{SiO}_{2}$ (20 nm thick)/Pt (5 nm thick)/Py (10 nm thick) quadrilayer film. Although a spin current is interrupted across the YIG/Pt interface, voltage anomalies persist near $\theta_{H}=\pm20$ deg.. Moreover, the quadrilayer film reveals the similar $\theta_H$ dependence of $V_{\mathrm{SYM}}$ at FMR of YIG as observed in the trilayer film [see Fig. 4(a)]. The persistence and the angular dependence both suggest that a dynamic exchange coupling is irrelevant to the trilayer film. 

The voltage anomalies in the YIG/$\mathrm{SiO}_{2}$/Pt/Py quadrilayer film suggest that unconventional voltage originates in spin pumping at the Py/Pt interface. To reinforce this argument, we prepared a YIG/Pt (5 nm thick)/$\mathrm{SiO}_{2}$ (20 nm thick)/Py (10 nm thick) quadrilayer film, where the $\mathrm{SiO}_{2}$ layer suppresses a dynamic exchange coupling by interrupting a spin current across the Py/Pt interface. The unconventional voltage is found to disappear [see Fig. 4(b)]. This result further proves that unconventional voltage originates in spin pumping at the Py/Pt interface.

We need to explain at least the following four features of the unconventional voltage: (feature 1) the unconventional voltage persists even when the ferromagnets are separated by up to about 20 nm; (feature 2) the voltage originates in spin pumping at the Py/Pt interface; (feature 3) the voltage appears when both the ferromagnets are excited simultaneously; (feature 4) the voltage takes extrema at $\theta_{H}$ where YIG and Py are equal in $H_{\mathrm{FMR}}$. 

The long-range nature (feature 1) suggests that a dipolar coupling may be relevant to the unconventional voltage. As shown below, however, this coupling alone is unlikely to yield our results consistently. Magnetization dynamics in the YIG layer yields a time-dependent dipole field\cite{Dmytriiev:2010,Schafer:2012}. A sum of the time-dependent dipole field and a microwave field act on magnetization in the Py layer. The increased driving field may enhance the magnetization dynamics in the Py layer. Note that YIG and Py are greatly different in volume, the saturation magnetization, and the damping constant. These differences may enhance spin pumping of Py while having little influence on that of YIG. On the other hand, a static dipole coupling should lead to a resonance-field shift in the YIG/Pt/Py trilayer film. This shift will become prominent especially when YIG and Py are excited simultaneously.\cite{Heinrich:1990} At least we did not  observe resonance-field shifts greater than an error range of 10 mT. It is unreasonable to infer that a dynamic dipole field enhances magnetization dynamics while a static counterpart has little influence on resonance fields. The unconventional voltage will need contributing factors other than a dipolar coupling.


\section{Conclusion}
We investigated spin pumping in a YIG/Pt/Py trilayer film by means of ISHE. Unconventional voltage appears when both the ferromagnets are excited simultaneously by a microwave. The voltage is attributed to spin pumping at the Py/Pt interface. The voltage persists when a dynamic exchange coupling is suppressed. Several features are pointed out to characterize the unconventional voltage. The voltage persists with the ferromagnets separated by up to 20 nm (long-range feature). The voltage magnitude increases when YIG and Py are equal in $H_{\mathrm{FMR}}$ (dynamic feature). Considering these features, a time-dependent dipolar coupling was proposed to explain the origin of the unconventional voltage. However, our experiments show that a static dipolar coupling does not shift resonance fields. This result appears inconsistent with the proposed mechanism. The origin of the unconventional voltage requires further elucidation.  

\section*{ACKNOWLEDGEMENTS}

This work was supported by CREST-JST "Creaction of Nanosystems with Novel Functions through Process Integration", a Grant-in-Aid for Scientific Research (A) (24244051) from MEXT, Japan, The Murata Science Foundation, The Funding Program for Next Generation World-Leading Researchers, The Asahi Glass Foundation, The Noguchi Institute, and The Mitsubishi Foundation.

\newpage{}
\begin{figure}[h]
  \centering
    \includegraphics{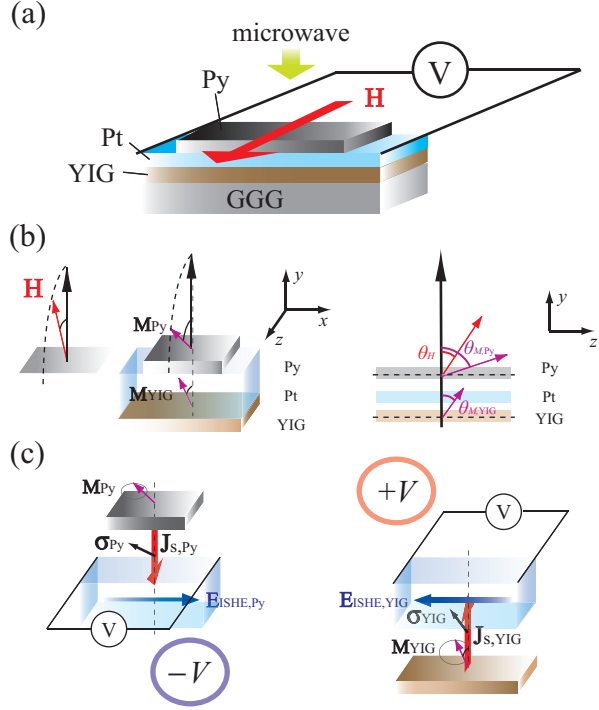}
  \caption{(a) A schematic illustration of the YIG/Pt/Py trilayer film. $\mathbf{H}$ is a static magnetic field. (b) Definitions of the magnetic-field angle $\theta_{H}$ and the magnetization angle $\theta_{M}$. $\mathbf{H}$ is a static magnetic field and $\mathbf{M}$ is magnetization. (c) A schematic illustration of spin pumping at the Py/Pt (left) and YIG/Pt interfaces (right). $\mathbf{M}$ denotes magnetization; $\mathbf{J}_{\mathrm{s}}$ the spatial direction of a spin current; $\mathbf{\boldsymbol\sigma}$ the polarization vector; $\mathbf{E}_{\mathrm{ISHE}}$ an ISHE-induced electric field.}
  \label{Fig.1.}
\end{figure}

\newpage{}
\begin{figure}[h]
  \centering
  \includegraphics{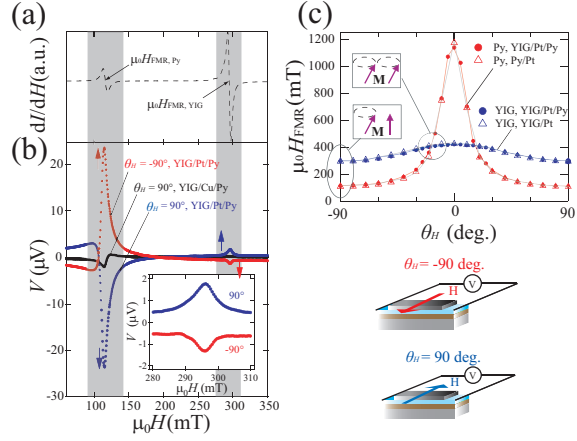}
  \caption{(a) The in-plane field $H$ dependence of the microwave abosrption signal $\mathrm{d}I/\mathrm{d}H$. $H_{\mathrm{FMR}}$ is a resonance field. (b) The in-plane field $H$ dependence of DC voltage $V$ for the YIG/Pt/Py trilayer film (red and blue circles) and the YIG/Cu/Py trilayer film (black circles). The inset shows the $H$ dependence of $V$ near FMR of YIG. (c) The field-angle $\theta_{H}$ dependence of the resonance field $H_{\mathrm{FMR}}$. The filled circles represent data for the YIG/Pt/Py trilayer film; the open triangles for the YIG/Pt and Py/Pt bilayer films. Data are blue for YIG and red for Py.}
  \label{Fig.2}
\end{figure}

\newpage{}
\begin{figure}[h]
  \centering
  \includegraphics{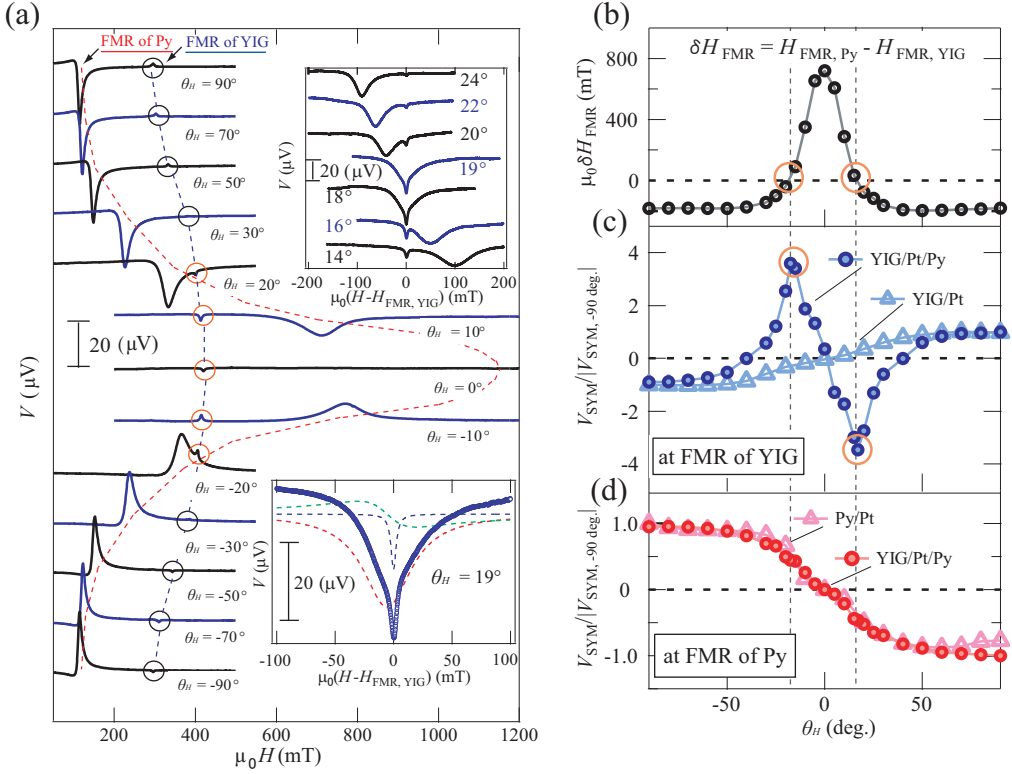}
  \caption{(a) The field $H$ dependence of DC voltage $V$ for the YIG/Pt/Py trilayer film at magnetic-field angles $\theta_{H}$ between $-90$ and $90$ deg. The blue (red) dashed line represents the resonance field of YIG (Py). Orange (black) circles indicate unconventional (conventional) voltage at FMR of YIG. The upper inset shows the $H$ dependence of $V$ near FMR of YIG. The lower inset shows experimental data (open circles) and fitting (dashed lines) at $\theta_{H} = 19$ deg. Red and blue dashed lines are a symmetric Lorenzian function and the green dashed line is an asymmetric Lorenzian function. (b) The magnetic-field angle $\theta_{H}$ dependence of the resonance-field difference $\delta H_{\mathrm{FMR}}$. $\delta H_{\mathrm{FMR}}$ is defined as $\delta H_{\mathrm{FMR}} = H_{\mathrm{FMR, Py}} - H_{\mathrm{FMR, YIG}}$. Orange circles indicate where YIG and Py are equal in $H_{\mathrm{FMR}}$. (c) The magnetic-field angle $\theta_{H}$ dependence of the symmetric component of DC voltage $V_{\mathrm{SYM}}$ at FMR of YIG. $V_{\mathrm{SYM}}/\left|V_{\mathrm{SYM}, -90\mathrm{deg}}\right|$ is the normalized DC voltage. Circles represent data for the YIG/Pt/Py trilayer film; triangles for the YIG/Pt bilayer film. Orange circles indicate where YIG and Py are equal in $H_{FMR}$. (d) The magnetic-field angle $\theta_{H}$ dependence of the symmetric component of DC voltage $V_{\mathrm{SYM}}$ at FMR of Py. $V_{\mathrm{SYM}}/\left|V_{\mathrm{SYM}, -90\mathrm{deg}}\right|$ is the normalized DC voltage. Circles represent data for the YIG/Pt/Py trilayer film; triangles for the YIG/Pt bilayer film.}
  \label{Fig.3.}
\end{figure}

\newpage{}
\begin{figure}[h]
  \centering
  \includegraphics{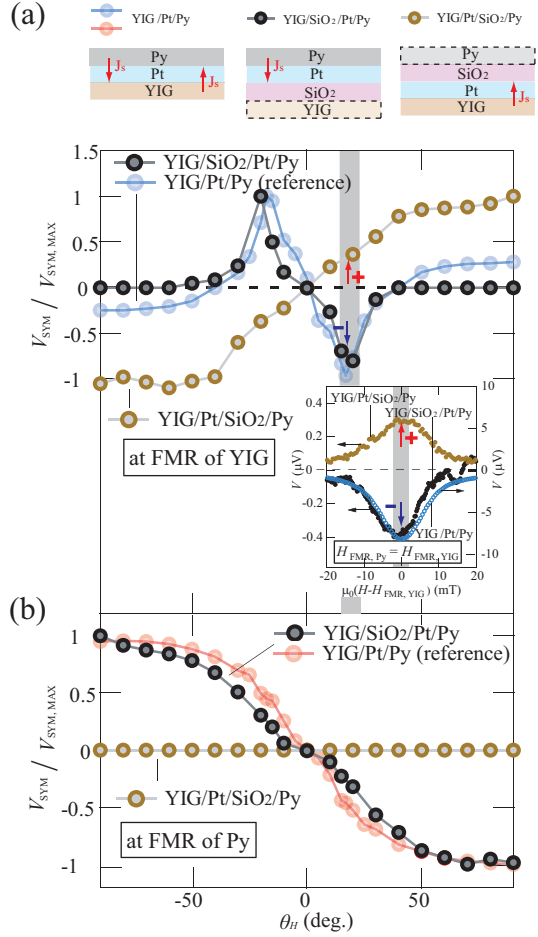}
  \caption{(a) The magnetic-field angle $\theta_{H}$ dependence of the symmetric component of DC voltage $V_{\mathrm{SYM}}$ at FMR of YIG. The black circles represent data for the YIG/$\mathrm{SiO}_2$/Pt/Py quadrilayer film; the dark brown circles for the YIG/Pt/$\mathrm{SiO}_2$/Py quadrilayer films; the sky blue circles for the YIG/Pt/Py trilayer film. The inset shows the magnetic field $H$ dependence of $V_{\mathrm{SYM}}$ near FMR of YIG. $\theta_{H}$ is such that YIG and Py are equal in the resonance field $H_{\mathrm{FMR}}$. The red (blue) arrow stresses a positive (negative) sign of $V_{\mathrm{SYM}}$. (b) The magnetic-field angle $\theta_{H}$ dependence of the symmetric component of DC voltage $V_{\mathrm{SYM}}$ at FMR of Py. The black circles represent data for the YIG/$\mathrm{SiO}_2$/Pt/Py quadrilayer film; the dark brown circles for the YIG/Pt/$\mathrm{SiO}_2$/Py quadrilayer films; the light pink circles for the YIG/Pt/Py trilayer film.}
  \label{Fig.4.}
\end{figure}

\newpage{}


\begin{thebibliography}{10}

\bibitem{Zutic:2004vi} Zutic, J. Fabian, and S. Das Sarma, Rev. Mod.
Phys. \textbf{76}, 323 (2004).

\bibitem{Fert:2008} A. Fert, Rev. Mod. Phys. \textbf{80}, 1517 (2008).

\bibitem{Slonczewski:1996} J. C. Slonczewski, J. Magn. Magn. Mater. \textbf{159}, L1 (1996).

\bibitem{Berger:1996} L. Berger, Phys. Rev. B \textbf{54}, 9353 (1996).

\bibitem{Kiselev:2003} S. I. Kiselev, J. C. Sankey, I. N. Krivorotov, N. C. Emley, R. J. Schoelkopf, R. A. Buhrman, and D. C. Ralph, Nature \textbf{425}, 380 (2003).

\bibitem{Mizukami:2002tt} S. Mizukami, Y. Ando, and T. Miyazaki,
Phys. Rev. B \textbf{66}, 104413 (2002).

\bibitem{Tserkovnyak:2005fr} Y. Tserkovnyak, A. Brataas, G. E. W.
Bauer, and B. I. Halperin, Rev. Mod. Phys. \textbf{77}, 1375 (2005).

\bibitem{Azevedo:2005} A. Azevedo, L. H. Vilela Le$\tilde{\mathrm{a}}$o, R. L. Rodriguez-Suarez, A. B. Oliveira, and S. M. Rezende, J. Appl. Phys. \textbf{97}, 10C715 (2005).

\bibitem{Kimura:2007} T. Kimura, Y. Otani, T. Sato, S. Takahashi, and S. Maekawa, Phys. Rev. Lett. \textbf{98}, 156601 (2007).

\bibitem{Saitoh:2006} E. Saitoh, M. Ueda, H.Miyajima, and G. Tatara, Appl. Phys. Lett. \textbf{88}, 182509 (2006).

\bibitem{Valenzuela:2006} S. O. Valenzuela and M. Tinkham, Nature \textbf{442}, 176 (2006).

\bibitem{Seki:2008} T. Seki, Y. Hasegawa, S. Mitani, S. Takahashi, H. Imamura, S. Maekawa, J. Nitta, and K. Takanashi, Nat. Mater. \textbf{7}, 125 (2008).

\bibitem{Takahashi:2002} S. Takahashi and S. Maekawa, Phys. Rev. Lett. \textbf{88}, 116601 (2002).

\bibitem{Ando:2008} K. Ando, Y. Kajiwara, S. Takahashi, S. Maekawa, K. Takemoto, M. Takatsu, and E. Saitoh, Phys. Rev. B \textbf{78}, 014413 (2008).

\bibitem{Ishibashi:2005} T. Ishibashi, A. Mizusawa, M. Nagai, S. Shimizu, K. Sato, N. Togashi, T. Mogi, M. Houchido, H. Sato, and K. Kuriyama, J. Appl. Phys. \textbf{97}, 013516 (2005).

\bibitem{Heinrich:2003} B. Heinrich, Y. Tserkovnyak, G. Woltersdorf, A. Brataas, R. Urban, and G. E. W. Bauer, Phys. Rev. Lett. \textbf{90}, 187601 (2003).

\bibitem{Ando:2011} K. Ando, S. Takahashi, J. Ieda, Y. Kajiwara, H. Nakayama, T. Yoshino, K. Harii, Y, Fujikawa, M. Matsuo, S. Maekawa, and E. Saitoh, J. Appl. Phys. \textbf{109}, 103913 (2011).


\bibitem{Chen:2013} L. Chen, F. Matsukura, and H. Ohno, Nat. Commun. \textbf{4}, 2055 (2013)


\bibitem{Uchida:2010} K. Uchida, H. Adachi, T. Ota, H. Nakayama, S. Maekawa, and E. Saitoh, Appl. Phys. Lett. \textbf{97}, 172505 (2010).

\bibitem{Dmytriiev:2010} O. Dmytriiev, T. Meitzler, E. Bankowski, A. Slavin, and V. Tiberkevich, J. Phys. Condens. Matter \textbf{22}, 136001 (2010).

\bibitem{Schafer:2012} S. Schafer, N. Pachauri, C. K. A. Mewes, T. Mewes, C. Kaiser, Q. Leng, and M. Pakala, Appl. Phys. Lett. \textbf{100}, 032402 (2012).

\bibitem{Heinrich:1990} B. Heinrich, Z. Celinski, J. F. Cochran, W. B. Muir, J. Rudd, Q. M. Zhong, A. S. Arrott, K. Myrtle, and J. Kirschner, Phys. Rev. Lett. \textbf{64}, 673 (1990) 





\end{thebibliography}
\end{document}